\documentclass[prb,preprint,showpacs,preprintnumbers,amsmath,amssymb]{revtex4}
\usepackage{graphicx}
\usepackage{dcolumn}
\usepackage{bm}

\begin{document}
\draft
\title{ Coexistence of antiferromagnetic order and unconventional superconductivity in  heavy fermion compounds CeRh$_{1-x}$Ir$_x$In$_5$: nuclear quadrupole resonance studies}
\author{Guo-qing Zheng$^{1,*}$, N. Yamaguchi$^1$, H. Kan$^1$,  Y. Kitaoka$^1$, J. L. Sarrao$^2$, P.G. Pagliuso$^2$, N.O. Moreno$^2$, J. D. Thompson$^2$}
\address{$^1$Department of Physical Science, Graduate School of Engineering
Science, Osaka University, Osaka 560-8531, Japan}
\address{$^2$ Condensed Matter and Thermal Physics, MS K764, Los Alamos National Laboratory, Los Alamos, NM 87545, USA}


\begin{abstract}
We present a systematic $^{115}$In NQR study on the heavy fermion
compounds CeRh$_{1-x}$Ir$_x$In$_5$ ($x$=0.25, 0.35, 0.45, 0.5,
0.55 and 0.75). The results provide strong evidence for the
microscopic coexistence of antiferromagnetic (AF) order and
superconductivity (SC) in the range of 0.35 $\leq x \leq$ 0.55. Specifically,
for $x$=0.5, $T_N$ is observed at 3 K with a subsequent onset of
superconductivity at $T_c$=0.9 K. 
$T_c$ reaches a maximum (0.94 K) at $x$=0.45 where $T_N$ is found to be the highest (4.0 K). 
Detailed analysis of the
measured spectra indicate that the same electrons participate in
both SC and AF order. The nuclear spin-lattice relaxation rate
$1/T_1$ shows a broad peak at $T_N$ and follows a $T^3$ variation
below $T_c$, the latter property indicating unconventional SC as
in CeIrIn$_5$ ($T_c$=0.4 K). We further find that, in the
coexistence region, the $T^3$ dependence of $1/T_1$ is replaced by
a $T$-linear variation below $T\sim$0.4 K, with the value
$\frac{(T_1)_{T_c}}{(T_1)_{low-T}}$ increasing with decreasing
$x$, likely due to low-lying magnetic excitations associated with
the coexisting magnetism.

\end{abstract}

\maketitle
\widetext

\section{Introduction}

Superconductivity and long-range magnetic order are two
outstanding quantum phenomena; however these ground states are not
generally displayed by the same electrons simultaneously. This is
because an internal magnetic field arising from magnetic order
usually destroys superconductivity. In the 1970s, a number of
materials were found to host both superconductivity and magnetic
order, but the two orders were due to different electrons and
occurred in spatially-separated regions \cite{Maple&Fisher}. This
is also true in the recently reported ruthenate-cuprate hybrid
compound RuSr$_2$RCu$_2$O$_8$ (R=rare earth) where the RuO and
CuO$_2$ planes are responsible for the magnetic order and
superconductivity, respectively \cite{Barnhart}. An exceptional
case is the heavy fermion compound UPd$_2$Al$_3$  in which
magnetic order and superconductivity coexist  homogeneously
\cite{Geibel,Tou}. In this system, however, it is believed that  the
multiple bands of uranium (U) electrons make such coexistence
possible. Namely, among three U-5f electrons, the two with
localized character are responsible for the magnetism and the
remaining one is responsible for superconductivity
\cite{Sato,Miyake}. Such "duality" may also be at work  in other
U-based heavy fermion magneto-superconductors
\cite{Geibel2,UGe2,URhGe}. It is therefore an outstanding question
whether   magnetic order and superconductivity due to the same
electrons  can coexist on a microscopic length scale.  Although it
has been proposed theoretically that magnetism and
superconductivity may be viewed as two sub-components of a unified
group and that they may coexist under certain conditions
\cite{Zhang}, accumulation of convincing experimental evidence is
important. The Ce-based heavy fermion compounds and high
superconducting transition-temperature ($T_c$) copper oxides  are
hosts of single-band magnetism or/and superconductivity, and are
therefore good candidate materials for exploring this problem.

Recently, it has been suggested that in the layered heavy fermion
compounds Ce(Rh$_{1-x}$Ir$_{x}$)In$_5$ \cite{Pagliuso} and
Ce(Rh$_{1-x}$Co$_{x}$)In$_5$ \cite{Maple} and also CeRhIn$_5$
under pressure \cite{Mito,Kawasaki}, antiferromagnetism and
superconductivity coexist. CeRh(Ir)In$_5$ crystallizes in a
tetragonal structure which consists of CeIn$_3$ layers separated
by a Rh(Ir)In$_2$ block. CeRhIn$_5$ is an antiferromagnet with
$T_N$=3.7 K,  but becomes superconducting under pressures above
1.6 GPa \cite{Hegger}. CeIrIn$_5$ is a superconductor at ambient
pressure with $T_c$=0.4 K \cite{Petrovic} and line nodes in the
superconducting energy gap \cite{Zheng}. It is remarkable that the
magnetic fluctuations exhibit  quasi two-dimensional character as
revealed by NQR \cite{Zheng} and neutron scattering \cite{Bao1}
measurements, probably reflecting the layered crystal structure.
Upon substituting Rh with Ir, superconductivity was found in
Ce(Rh$_{1-x}$Ir$_{x}$)In$_5$  for $x$$>$0.3, while magnetic order
continued to be observed around 3.8 K in the specific heat for
x$\leq$0.5 (Ref. \cite{Pagliuso}) and an internal magnetic field
was detected by muon spin rotation measurement \cite{Morris}.

In this paper, we present results obtained from nuclear quadrupole
resonance (NQR) measurements on Ce(Rh$_{1-x}$Ir$_{x}$)In$_5$ that
strongly suggest  that antiferromagnetic (AF) order coexists
microscopically with unconventional superconductivity (SC). We
find that upon replacing Rh with Ir in the antiferromagnet
CeRhIn$_5$,   the Neel temperature $T_N$ increases slightly with
increasing Ir content up to $x$=0.45 then decreases rapidly.
Superconductivity sets in above $x\sim$0.35 and $T_c$ reaches a
maximum of 0.94 K at $x$=0.45. The nuclear spin-lattice relaxation
rate $1/T_1$ shows a broad peak at $T_N$  and follows a $T^3$
variation below $T_c$, the latter feature  indicating that the SC
is unconventional  as in CeIrIn$_5$. In the coexistence region,
$1/T_1$ becomes proportional to $T$ at very low temperatures in
the superconducting state and  the value $T_1(T=T_c)/T_1$
increases in the order of x=0.55, 0.5 and 0.45, which suggests the
existence of low-lying magnetic excitations in addition to the
residual density of states (DOS) due to the presence of disorder.

The rest of the paper is organized as follows. The experimental
details are described in Section II.  In Section III, the NQR
spectroscopy that indicates the homogeneous alloying of the
samples is presented.  The results of the nuclear spin lattice
relaxation that evidence the coexistence of antiferromagnetism and
superconductivity are also presented in Section III, along with
evidence for the unconventional nature of the superconductivity.
We conclude in Section IV, following a brief discussion of the
phase diagram deduced from our NQR measurements.

\section{Experimetal}

Single crystals of Ce(Rh$_{1-x}$Ir$_x$)In$_5$ used in this study
were grown by the In-flux method \cite{Hegger}.  For NQR
measurements, the single crystals were crushed into  a powder of
moderate particle size to allow  maximal penetration of the
oscillating magnetic field, $H_1$, used in the NQR measurements.
The measurements below 1.4 K were performed by using a
$^{3}$He/$^{4}$He dilution refrigerator. NQR experiments were
performed using a home-built phase-coherent spectrometer. A
standard $\pi$/2-$\pi$-echo pulse sequence was used. A small $H_1$
was used to avoid possible heating by the RF pulse below 1 K; the
$\pi$/2 pulse length is about 20 micro-seconds. A
CuBe piston-cylinder device \cite{ZhengSSC}, filled with Si-based organic liquid
as a pressure-transmitting medium, was used to generate high
pressure. The NQR coil was put inside a Teflon cell. To calibrate the pressure  at low temperatures,
the reduction in $T_{\rm c}$ of Sn metal under pressure was
monitored by resistivity measurements \cite{smith}.  $T_c$ of the
samples was determined from the ac susceptibility measured by
using the NQR coil at a frequency of $\sim$32 MHz, and from the
$T_1$ data (see below). $1/T_1$  was measured by the
saturation-recovery method. The value of $1/T_1$ was unambiguously
extracted from a good fitting of the nuclear magnetization to the
expected theoretical curve \cite{Maclaughlin,Roos} (discussed in
detail below).

\begin{figure}
\begin{center}
\includegraphics[scale=0.5]{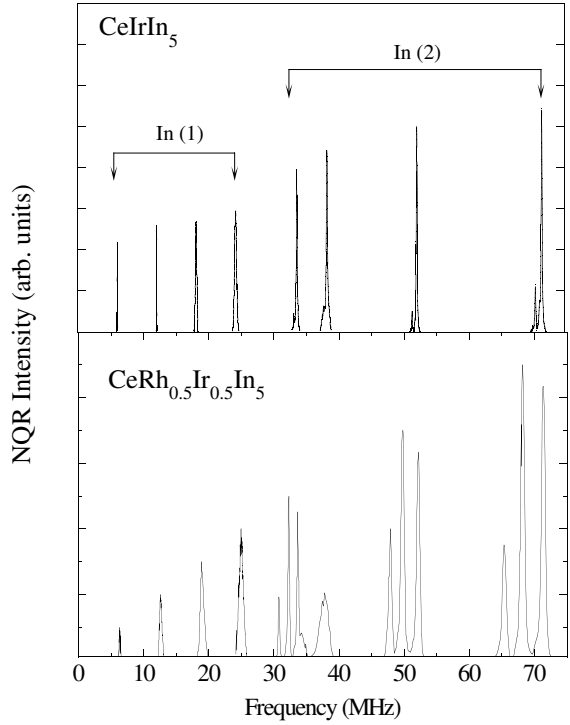}
\caption{ $^{115}$In NQR spectra at $T$=4.2 K for CeIrIn$_5$ (upper panel \cite{Zheng}), and for CeRh$_{0.5}$Ir$_{0.5}$In$_5$ (lower panel).}
\label{fig:1}
\end{center}
\end{figure}

\section{Results and discussion}

\subsection{Evidence for homogeneous alloying from NQR spectra}

There are two inequivalent crystallographic sites of In in
Ce(Rh$_{1-x}$Ir$_{x}$)In$_5$: the In(1) site in the CeIn$_{3}$
plane and the In(2) site in the Rh(Ir)In$_2$ block.  The NQR
spectra for the In(1) site consist of four  equally-spaced
transition lines separated by $\nu_Q$, while the In(2) spectra are
composed of four un-equally separated lines between 30 and 72 MHz.
The spectra of CeIrIn$_5$ (Ref. \cite{Zheng}) is reproduced in
Fig. 1(a).  Here $\nu_Q$ is defined as the parameter in the
following Hamiltonian,
\begin{eqnarray}
H_Q = \frac{h\nu_Q}{6}(3I_z^2-I(I+1)+\frac{1}{2}\eta (I_{+}^2+I_{-}^2))
\end{eqnarray}

where
\begin{eqnarray}
\nu_Q = \frac{eQV_{zz}}{6I(2I+1)}
\end{eqnarray}
and
\begin{eqnarray}
\eta = \frac{V_{xx}-V_{yy}}{V_{zz}}
\end{eqnarray}

\begin{figure}
\begin{center}
\includegraphics[scale=0.4]{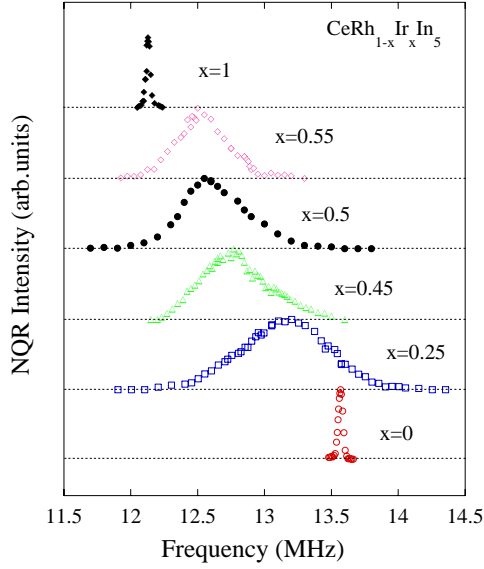}
\caption{ $^{115}$In NQR line shape ($\pm 3/2 \leftrightarrow \pm$5/2 transition) of the In(1) site in CeRh$_{1-x}$Ir$_x$In$_{5}$ at $T$=4.2 K for various Ir content. The horizontal line under each spectrum is the position of zero intensity for that spectrum.}
\label{fig:2}
\end{center}
\end{figure}

A representative spectra for CeRh$_{0.5}$Ir$_{0.5}$In$_5$ is shown
in Figure 1(b) . Two effects due to alloying are readily seen in
this spectra. First, the transition lines for In(1) are broadened.
Second, each transition for In(2) is split into three lines.
Although naively this behavior might suggest phase segregation, we
argue below by inspecting the Ir-concentration dependence of the
spectra, that there is no phase separation in the alloyed sample;
rather the sample is globally homogeneous.

Figure 2 shows the NQR line shape at T=4.2 K of the 2$\nu_Q$
transition at the In(1) site for various Ir contents. The $\nu_Q$
decreases monotonically  from 6.78 MHz ($x$=0)
\cite{Curro} to 6.065 MHz ($x$=1) \cite{Zheng}, suggesting a
smooth evolution of the lattice upon alloying, in agreement with
x-ray diffraction measurements \cite{Pagliuso}.  It should be
emphasized that no trace of pure CeRhIn$_5$ or CeIrIn$_5$ is found
in the alloyed samples because no peaks corresponding to $x$=0 or
$x$=1 were observed.

\begin{figure}
\begin{center}
\includegraphics[scale=0.5]{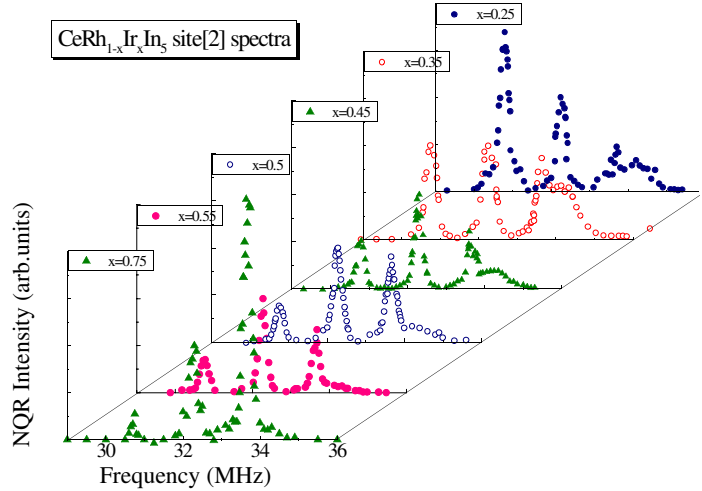}
\caption{ $^{115}$In NQR line shape ($\pm 3/2 \leftrightarrow \pm$5/2 transition) of the In(2) site of CeRh$_{1-x}$Ir$_x$In$_{5}$ at $T$=4.2 K for various Ir content. In this plot, the vertical axis was adjusted so that all samples have the same height for the central peak (around 32.2 MHz). The signal around 35 MHz for low $x$ is from the second lowest transition ($\pm 1/2 \leftrightarrow \pm$3/2 transition); also see Fig. 1(b).}
\label{fig:3}
\end{center}
\end{figure}

Figure 3 shows the spectra corresponding to the lowest  transition
($m=\pm 3/2 \leftrightarrow \pm 5/2$) line of the In(2) site for
various Ir concentration ranging from $x$=0.25 to 0.75. It is
interesting that the  positions of the three peaks do not change
with Ir concentration (Fig. 4(a)), but the relative intensity
distribution among these lines does (Fig. 4(b)). Also, the left
peak is at the same position of the $m=\pm 3/2 \leftrightarrow \pm
5/2$ transition for CeRhIn$_{5}$, while the right peak is at the
same position as the corresponding transition for pure CeIrIn$_5$.
The central peak is characterized by $\nu_{Q}$=17.37 MHz and
$\eta$=0.473.

\begin{figure}
\begin{center}
\includegraphics[scale=0.5]{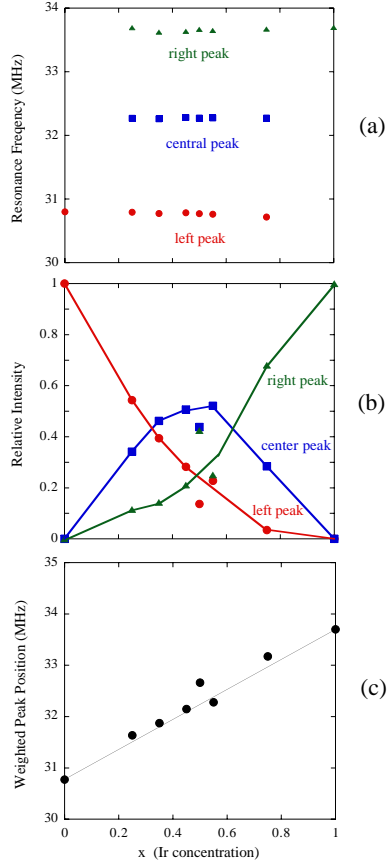}
\caption{(a)  Peak frequency of the three $\pm 3/2 \leftrightarrow \pm$5/2 transition lines of the In(2) site of CeRh$_{1-x}$Ir$_x$In$_{5}$ at $T$=4.2 K. (b) Ir-content dependence of the relative intensity of the three $\pm 3/2 \leftrightarrow \pm$5/2 transition lines of the In(2) site of CeRh$_{1-x}$Ir$_x$In$_{5}$. (c) Ir-content dependence of the peak frequency multiplied by the relative intensity for the three In(2)  $\pm 3/2 \leftrightarrow \pm$5/2 transition lines.}
\label{fig:4}
\end{center}
\end{figure}

Figure 4(c) depicts a quantity that is the relative intensity
shown in Fig. 4(b) multiplied by  the corresponding peak position
shown in Fig. 4(a). Most simply, this corresponds to the "weighted
peak position" or "averaged resonance frequency" for the $m=\pm
3/2 \leftrightarrow \pm 5/2$ transition. Note that this quantity
increases smoothly with increasing Ir concentration.

The results shown in Fig. 4 can be interpreted as follows. In(2)
has two nearest neighbor $M$ (Rh, Ir) sites. There are $x$ Ir
atoms and $(1-x)$ Rh atoms for a given alloy concentration x. If
the NQR frequency is sensitive to the local environment, there
will be three resonance lines depending on the nearest neighbor
configuration of a given In(2), namely, (Rh, Rh), (Rh, Ir) or
(Ir,Ir).  The intensity of each peak will be proportional to the
probability that In(2) has a corresponding nearest neighbor pair,
namely, (Rh, Rh), (Rh, Ir) or (Ir,Ir). Figure 4 strongly suggests
that this is the case, with the central transition corresponding
to the case with (Rh,Ir) nearest neighbors.

In such a scenario, one might then wonder why In(1) only sees an
averaged environment. This is probably because the wave function
mixing between In(1) and the $M$ atom is weaker than in the case
of In(2), because In(1) is farther away from $M$. In addition,
In(1) has eight nearest neighbor $M$ atoms. The effect of having
different nearest-neighbor pair is thus further averaged out. As a
result, each In(1) transition is observed as a broadened line.
This is in contrast to the case of In(2) whose $p$-orbital
directly mixes with those of $M$. Since $\nu_Q$ is dominated by
the on-site electronic configuration \cite{Harima}, the stronger
coupling between In(2) and $M$ atoms gives rise to three distinct
resonance lines in the alloyed samples rather than a broad
'single' transition as in the case of In(1).

Although the In(2) transition is sensitive to the local atomic
configuration, it should be emphasized that globally the
electronic states are quite homogeneous, as evidenced by the
results of spin-lattice relaxation measurements described in the
next subsection.

\subsection{Nuclear spin lattice relaxation and the magnetic ordering}

The $1/T_1$  measurements were performed at the peak of the
2$\nu_Q$ transition  ($m=\pm 3/2 \leftrightarrow \pm 5/2$ for the
In(1) site and at the central peak of the three lowest frequency
transition ($m=\pm 3/2 \leftrightarrow \pm 5/2$) lines for the
In(2) site. Figure 5 shows the decay curve of the nuclear
magnetization for $x$=0.45 at three typical temperatures above and
below $T_N$ and $T_c$. At T=0.2K we used a small tipping-angle
pulse so that the magnetization is less saturated at small delay
time. The decay curve can be fitted by a single component of $T_1$
to the theoretical curve \cite{Maclaughlin},
\begin{eqnarray}
1-\frac{M(t)}{M_0} =  \frac{1}{33}
exp(-3\frac{t}{T_1})+\frac{20}{143}
exp(-10\frac{t}{T_1})+\frac{4}{165}exp(-21\frac{t}{T_1})+\frac{576}{715}exp(-36\frac{t}{T_1})
\end{eqnarray}
The same quality of data were obtained for all alloys and also for
the In(2) site, whose nuclear magnetization is fitted to the
theoretical curve \cite{Roos}with a single component of $T_1$.
\begin{eqnarray}
1-\frac{M(t)}{M_0} & =& 0.02421exp(-2.93355\frac{t}{T_1})+0.03961 exp(-8.30137\frac{t}{T_1}) \nonumber \\
& & +0.09771 exp(-16.30355\frac{t}{T_1})+0.83847 exp(-29.75056\frac{t}{T_1})
\end{eqnarray}

\begin{figure}
\begin{center}
\includegraphics[scale=0.5]{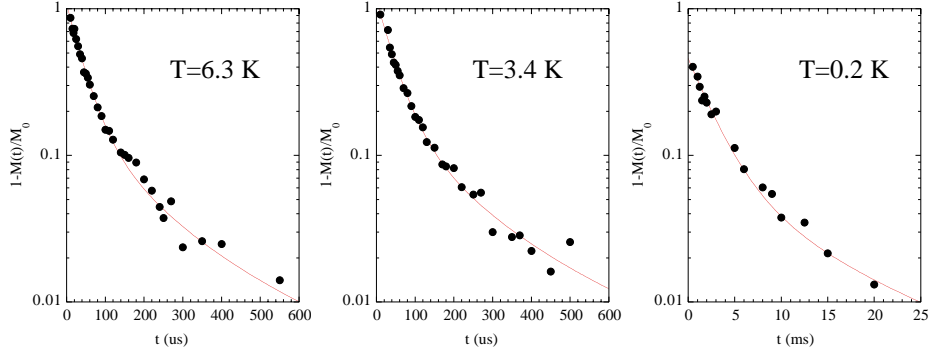}
\caption{ Time dependence of the nuclear magnetization of the In(1) site at various temperatures in CeRh$_{0.55}$Ir$_{0.45}$In$_5$. The curves are fitting to equation (4) in the text.}
\label{fig:5)}
\end{center}
\end{figure}

The successful fitting of the nuclear magnetization to the
theoretical curve with a single $T_1$ component is a good
indicator of the homogeneous nature of the electronic state.
Figure 6 shows the temperature dependence of $1/T_1$ measured at
the three peaks of In(2) for $x$=0.35. It can be seen that all
sites show a quite similar $T$ dependence. Namely, there is a peak
around $T$=4 K, although the peak height is reduced as compared to
$x$=0 \cite{Mito3}. The absolute value is also very similar. In
the figure, the origin for the left and right peaks  were shifted
for clarity. These results indicate that the three peaks probe the
same electronic state despite the fact that they arise from
different nearest-neighbor $M$ configurations.

\begin{figure}
\begin{center}
\includegraphics[scale=0.5]{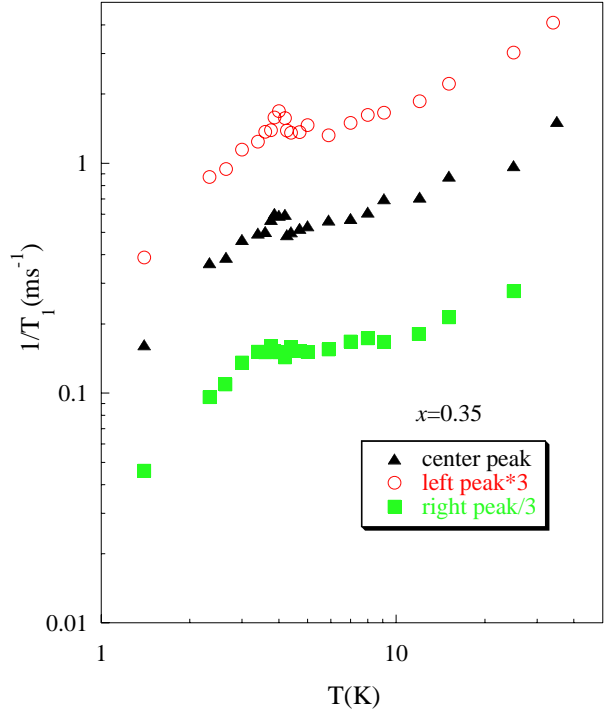}
\caption{Temperature dependence of $1/T_1$ at the three $\pm 3/2 \leftrightarrow \pm$5/2 transition lines of the In(2) site of CeRh$_{0.65}$Ir$_{0.35}$In$_{5}$.  For clarity, $1/T_1$ at the left peak was multiplied by 3, while that for the right peak was divided by 3.}
\label{fig:6)}
\end{center}
\end{figure}

Figure 7 shows the evolution of the $T$ dependence of $1/T_1$ at
the central In(2) transition for various Ir concentrations. It is
evident that the peak temperature and the peak height change with
the Ir concentration. We associate this peak with the Neel
ordering temperature, $T_N$,  at which $1/T_1$ increases  due to
critical slowing down. $T_N$ determined in this manner correspond
well with that inferred from the specific heat \cite{Pagliuso} and
$\mu$sr measurements \cite{Morris}. Interestingly, $T_N$ first
increases gradually with increasing Ir content up to $x$=0.45 then
decreases rapidly.  For $x$=0.5, $T_N$ is reduced to 3 K. For
$x$=0.55, no feature is seen in the $T$-dependence  of $1/T_1$
(for clarity of Fig. 7, data are not shown ), thus it becomes difficult to
identify $T_N$.

\begin{figure}
\begin{center}
\includegraphics[scale=0.5]{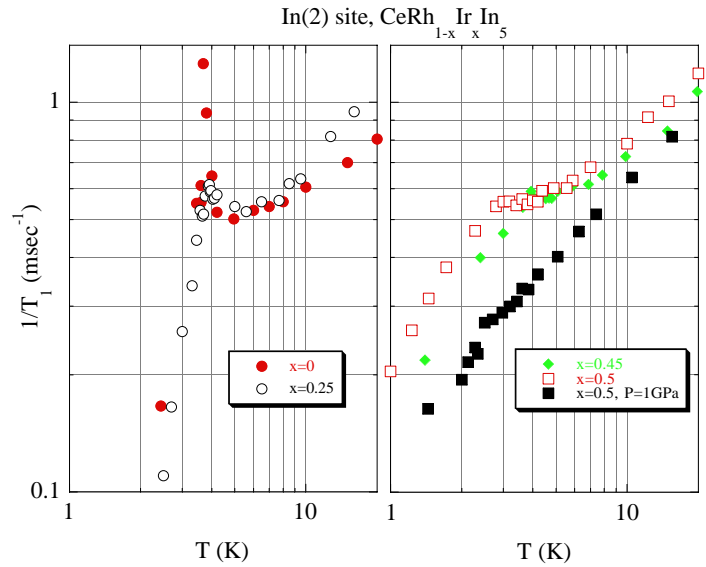}
\caption{Typical data sets of $1/T_1$  measured at the central peak of the In(2) $\pm 3/2 \leftrightarrow \pm$5/2 transition in CeRh$_{1-x}$Ir$_x$In$_{5}$.}
\label{fig:7)}
\end{center}
\end{figure}

$T_N$ inferred from the peak in $1/T_1$  is sensitive to
externally-applied hydrostatic pressure, as in pure CeRhIn$_5$. In
the right panel of Fig. 7 is shown the $T_1$ result under a
pressure of 1.02 GPa for the $x$=0.5 sample. The broad peak seen
at ambient pressure is suppressed, and instead a distinct decrease
of $1/T_1$ is found at 2.5 K, which resembles the case of pure
CeRhIn$_5$ in which the application of pressure reduces the height
of the peak at $T_N$ \cite{Mito3,Kawasaki2,Mito2,Kohori} and eventually
suppresses the peak under $P$=1.7 GPa \cite{Mito2}. Thus, as in
pure CeRhIn$_5$,   $T_1$ can serve as a probe to determine $T_N$.

Figure 8  shows typical data sets of $1/T_1$ measured at the In(1)
site. The anomaly at $T_N$ is also  visible at the In(1) site,
although it is less clear presumably because the peak at $T_N$ at
this site is already rather weak, even in the undoped compound.

\begin{figure}
\begin{center}
\includegraphics[scale=0.35]{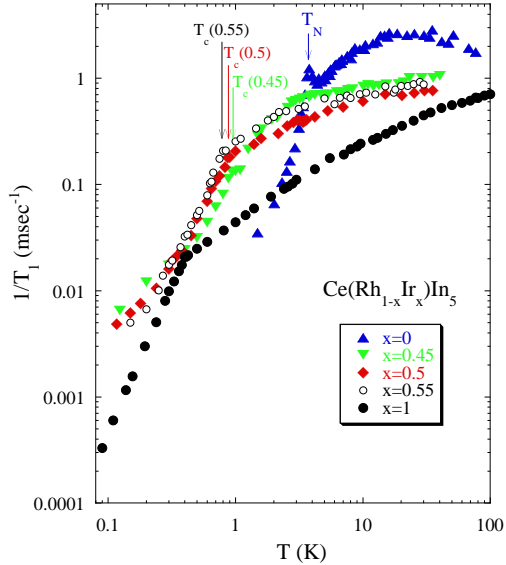}
\caption{Typical data sets of $^{115}$In  $1/T_1$    measured at the In(1) site of CeRh$_{1-x}$Ir$_x$In$_{5}$. Data for $x$=1 and 0 are from Ref. \cite{Zheng} and Ref.\cite{Mito3,Kawasaki2}, respectively. }
\label{fig:8)}
\end{center}
\end{figure}

The non-monotonic change of $T_N$ as a function of $x$ may be
attributed to the increase of exchange coupling between 4f spins
which is overcome by the increase of coupling between 4f spins and
conduction electrons above $x=0.45$,  as inferred from Doniach's
treatment of the Kondo necklace  \cite{Doniach}.  This result also
resembles the behavior of CeRhIn$_5$ \cite{Mito,Kawasaki} as a
function of pressure and indicates that the substitution of Ir for
Rh acts as chemical pressure in CeRhIn$_5$.

Due to the broadening of the spectra upon alloying, it is
difficult to estimate  precisely the internal magnetic field in
the ordered state. The Hamiltonion in the presence of magnetic
field is given by

\begin{eqnarray}
H = H_Q+H_{Zeeman}
\end{eqnarray}

where $H_Q$ is given by eq. (1) and
\begin{eqnarray}
H_{Zeeman} = - \gamma\hbar (H_xI_x+H_yI_y+H_zI_z)
\end{eqnarray}

\begin{figure}
\begin{center}
\includegraphics[scale=0.5]{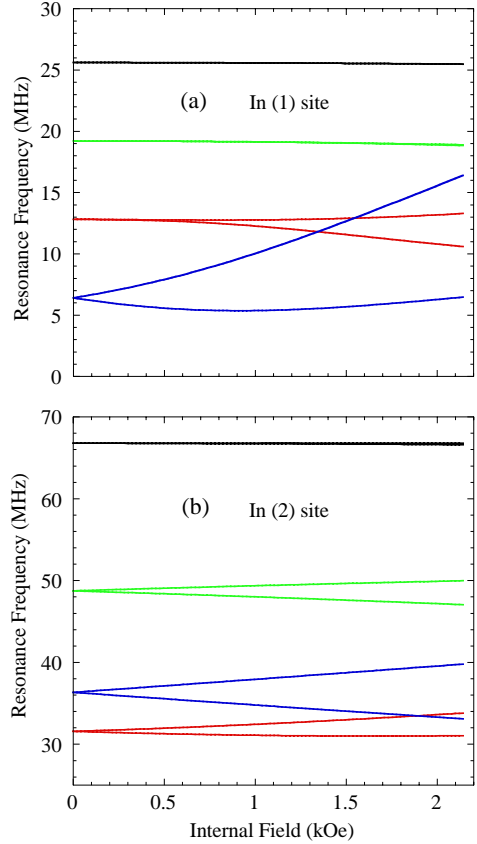}
\caption{Calculated evolution of the  NQR lines in the presence of internal field along the $a$-axis for In(1) site (a) and In(2) site (b).}
\label{fig:9(a))}
\end{center}
\end{figure}

In the present case, $V_{zz}$ is along the crystal c-axis.
Assuming an internal magnetic field in the ab-plane, which is the
case for CeRhIn$_{5}$, the evolution of the resonance frequency
for each transition is calculated for the In(1) site (Fig. 9(a))
and for the In(2) site (Fig. 9(b)). Here, the field is assumed to
be along x-direction. Note that even the $m=\pm 3/2
\leftrightarrow \pm 5/2$ transition for the In(2) site, which has
a  FWHM of 0.26 MHz and is the sharpest  among all transitions in
the alloyed samples, does not show an appreciable change between
$T$=4.2 K (above $T_N$) and  $T$=1.4 K (below $T_N$), see Fig. 10.
This suggests that the internal magnetic field at the In(2) site
is less than 200 Oe for $x$=0.5, as inferred from the expected
splitting deduced from Fig. 9. Such a small internal field, which is samller by a factor of 10 than that in CeRhIn$_5$ \cite {Curro},  could
be due to  a moderate reduction of the ordered moment \cite {Christianson}  with a concomitant
reduction of the hyperfine coupling \cite {Curro2}.

\begin{figure}
\begin{center}
\includegraphics[scale=0.5]{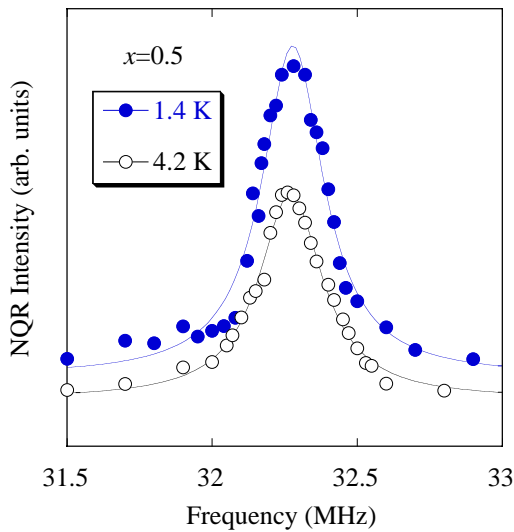}
\caption{The central peak of the $\pm 1/2 \leftrightarrow \pm$3/2 transition for CeRh$_{0.5}$Ir$_{0.5}$In$_5$ at $T$=4.2 K and 1.4 K. For clarity, the horizon has been shifted.}
\label{fig:10)}
\end{center}
\end{figure}

\subsection{Superconducting state}

Next, we  discuss the low temperature behavior of
Ce(Rh$_{1-x}$Ir$_{x}$)In$_5$ well below $T_N$. Figure 11 shows
$1/T_1$ for both the In(1) and In(2) sites at  low temperatures
for the $x$=0.5 sample. Below $T_c$=0.9 K,  $1/T_1$ decreases
sharply with no coherence peak, following a $T^3$ variation down
to $T$=0.45 K. The observation of the $T^3$ behavior is  strong
evidence for the existence of line nodes in the superconducting
gap function \cite{Zheng}. For an s-wave  gap, $1/T_1$ would show
a coherence peak just below $T_c$ followed by an exponential
decrease upon further decreasing $T$. Because $1/T_1$ is measured
at the same transition for the entire measured temperature range,
our results suggest that antiferromagnetic order and
superconductivity are due to the same electronic state derived
from the Ce-4f$^1$ electron. If the two ordered states occurred in
spatially-separated regions, the nuclear-magnetization decay curve
would have been composed of two components (two $T_1$'s) below
$T_N$, contradicting  the single-component decay curve we observe.
It is noteworthy that just above $T_c$, $1/T_1$ tends to be
proportional to $T$, which suggests that there remains a finite
density of states (DOS) at the Fermi level ($E_F$) in the
magnetically ordered state, since $1/T_1T$ is dominantly
proportional to the square of the low-energy  DOS at such low-$T$ (see below, eq. (9)).
This suggests that the gap opening due to the antiferromagnetic
order is incomplete, in contrast to the behavior observed in pure
CeRhIn$_5$ where the gap is more fully developed, leading to a
stronger decrease of $1/T_1$ (see Fig. 8). This remnant of some
part of the Fermi surface  may be important for superconductivity
to set in even in the magnetically ordered state.

\begin{figure}
\begin{center}
\includegraphics[scale=0.5]{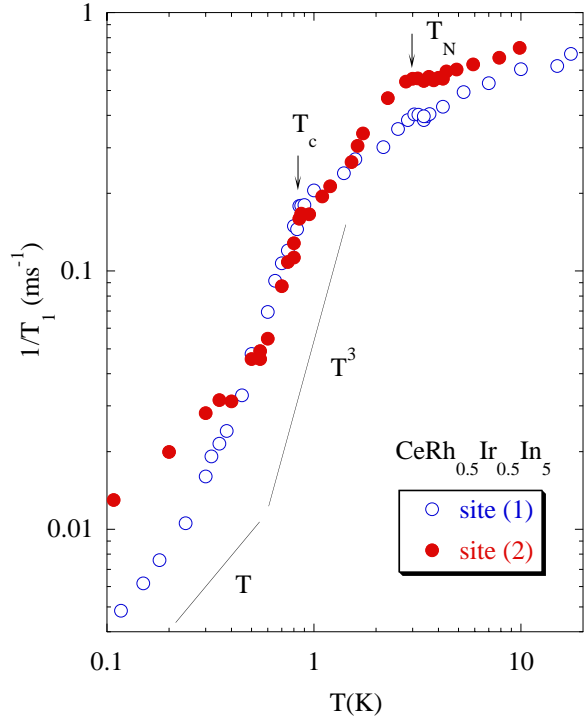}
\caption{The $1/T_1$ results at low temperatures for CeRh$_{0.5}$Ir$_{0.5}$In$_5$ measured at the In(1) at In(2) sites, respectively. The two solid lines indicate the $T^3$ and $T$-linear variations, respectively.}
\label{fig:11)}
\end{center}
\end{figure}

Finally, let us compare the superconducting behavior for $x$=0.45,
0.5 and 0.55. Figure 12 shows the ac-susceptibility (ac-$\chi$)
measured using our NQR coil. Although it is hard to determine the
onset temperature of the superconductivity from ac-$\chi$, it can
be seen that the mid-point of the transition increases in the
order of $x$=0.55, 0.5 and 0.45. $T_c$ determined from the point
at which $1/T_1$ displays a distinct drop is 0.8 K, 0.9 K and 0.94
K for $x$=0.55, 0.5 and 0.45, respectively. Figure 13 shows
$1/T_1$ normalized by its value at $T_c$ plotted against the
reduced temperature $T/T_c$ for $x$=0.55, 0.5 and 0.45. Just below
$T_c$, $1/T_1$ shows identical behavior for all samples, but at
lower temperatures strong variation is observed. In particular,
below $T\sim$ 0.4 K, $1/T_1$ becomes again  proportional to $T$,
and the normalized value of $1/T_1$ increases in the order
$x$=0.55, 0.5 and 0.45.

\begin{figure}
\begin{center}
\includegraphics[scale=0.5]{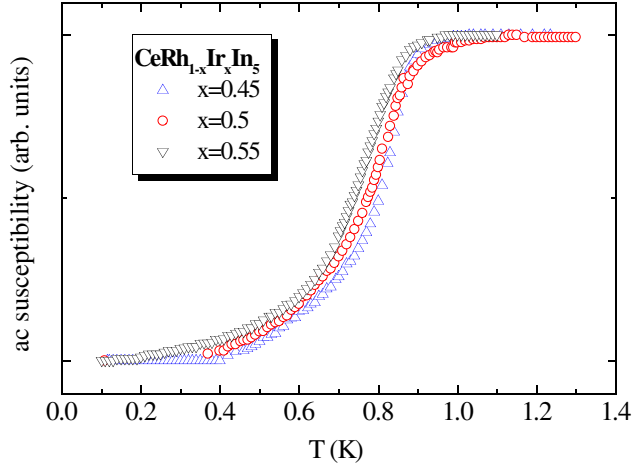}
\caption{The ac-susceptibility  for CeRh$_{1-x}$Ir$_{x}$In$_5$  ($x$=0.45, 0.5 and 0.55).}
\label{fig:12)}
\end{center}
\end{figure}

The most straightforward explanation for $T$-linear  $1/T_1$ at
low-$T$ would be  the presence of disorder that produces a  finite
DOS remaining at $E_F$. By assuming a gap function with line
nodes,
\begin{eqnarray}
\Delta(\theta)=\Delta_0 cos(\theta)
\end{eqnarray}
and with a finite residual DOS, $N_{res}$ (Ref.\cite{Miyake2}),
we tried to fit the data in the superconducting state to
\begin{eqnarray}
\frac{T_1(T=T_c)}{T_{1}}& = & \frac{2}{k_BT_c}\int (\frac{N_{s}(E)}{N_0})^{2}f(E)(1-f(E))dE,
\end{eqnarray}
where $\frac{N_{s}(E)}{N_{0}}=\frac{E}{\sqrt{E^{2}-\Delta^{2}}}$  with $N_{0}$ being the DOS in the normal state and $f(E)$ being the Fermi function.
 The resulting fitting parameters are $N_{res}/N_0$=0.32, 0.45 and 0.63 for $x$=0.55, 0.5 and 0.45, respectively, with $\Delta_0$=2.5$k_BT_c$ for all samples.
In such a case, however, one would expect  $N_{res}$ to be the
same for $x$=0.55 and 0.45, because the amount of disorder is
expected to be similar. The much larger $N_{res}$ inferred for $x$=0.45 than
$x$=0.55 suggests an additional mechanism. We propose that this
additional $N_{res}$ comes from low-lying magnetic excitations
associated with the coexisting magnetic ordering that is more well
developed at lower values of $x$ . Similar $N_{res}$ was seen in
CeRhIn$_5$ under a pressure of 1.6 GPa where magnetism also
coexists with superconductivity. In this case the observed
behavior was interpreted as due to a gapless $p$-wave
superconducting state \cite{Fuseya}, or due to additional nodes in the d-wave order parameter \cite{Bang}.

\begin{figure}
\begin{center}
\includegraphics[scale=0.5]{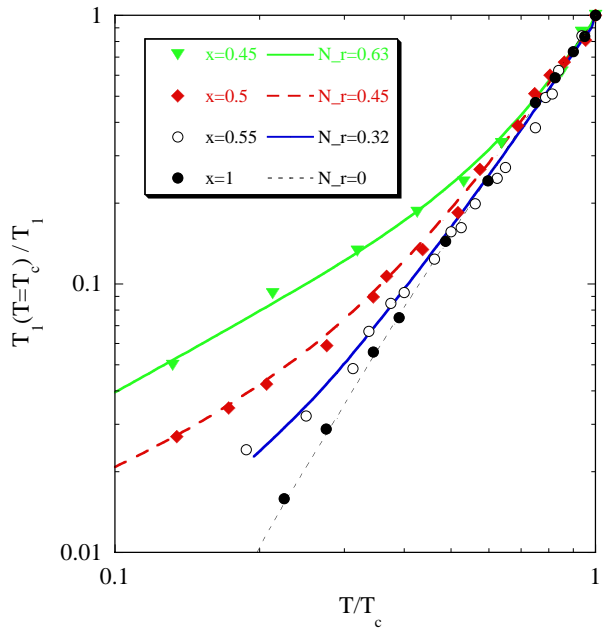}
\caption{The normalized $T_1$ plotted against the reduced temperature  for CeRh$_{1-x}$Ir$_{x}$In$_5$ at the In(1) site. The solid curves are fits to the data as described in the text. $N_r$ is for short of $N_{res}/N_0$.}
\label{fig:13)}
\end{center}
\end{figure}

On the other hand, the larger $N_{res}$ for the In(2) site than
for In(1) site may be due to a larger disorder contribution for
this site. This is because  the source of disorder in the present
case is in the Rh(Ir)In$_2$ block. The In(2) site is naturally
more sensitive to such disorder than the In(1) site which is
farther removed from this block. A similar case was seen  in
high-$T_c$ copper oxide superconductors. In
Tl$_2$Ba$_2$Ca$_2$Cu$_3$O$_{10}$ ($T_c$=117 K) \cite{Zheng1},
disorder due to inter-substitution of Ca/Tl occurs in the Ca
layer. As a consequence, the Cu(1) site sandwiched by two Ca
layers sees a larger  $N_{res}$ than the Cu(2) site which is
adjacent to only one of the Ca layers.

\begin{figure}
\begin{center}
\includegraphics[scale=0.5]{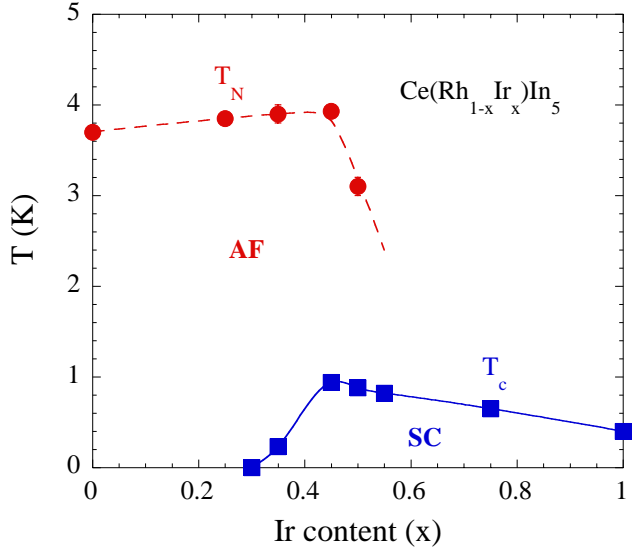}
\caption{The phase diagram of CeRh$_{1-x}$Ir$_x$In$_{5}$ obtained from NQR measurements. AF and SC mean antiferromagnetic and superconducting states, respectively.}
\label{fig:14)}
\end{center}
\end{figure}

\subsection{phase diagram}

The phase diagram shown in Fig. 14 summarizes our results. Upon
doping with Ir, the system undergoes a quantum phase transition
from an antiferromagnet ($x$=0) to a superconductor ($x$=1), with
an intervening region where antiferromagnetic and superconducting
orders  coexist. Our results show that this behavior, reported
previously based on thermodynamic data \cite{Pagliuso}, is
confirmed microscopically. $T_c$ reaches a maximum at $x$=0.45 ($T_c$=0.94 K),
while $T_N$ is found to be the highest ($T_N$=4.0 K). The enhancement of $T_c$ in the antiferromagnetically
ordered state is most interesting, suggesting  the importance of
magnetism in producing the superconductivity.
  Recently,  antiferromagnetism and superconductivity was found to coexist also in CeRhIn$_5$ under external pressures \cite{Mito,Kawasaki}, but the coexistent region is rather narrow there. More importantly, in the present case superconductivity develops well inside the ordered state and $T_c$ increases when approaching the maximum of $T_N$, whereas   $T_c$ reaches a maximum after $T_N$ disappears in hydrostatically-pressurized CeRhIn$_5$.
The  observed phase diagram may be understood in the framework of
SO(5) theory in which the 5-component super-spin can be rotated by
a chemical potential from the  subspace  of antiferromagnetic
order to the  subspace  of d-wave superconductivity and vice versa
\cite{Zhang}. However, a microscopic description of how the same 4f$^1$
electron can display both  magnetic order and superconductivity is
still lacking.

\section{Conclusion}

In conclusion, we have carried out an extensive $^{115}$In NQR
study on CeRh$_{1-x}$Ir$_x$In$_5$. We find that the substitution
of Ir for Rh in the antiferromagnet CeRhIn$_5$ acts as  chemical
pressure. With increasing Ir content ($x$), $T_N$ increases
slightly up to $x$=0.45, then decreases rapidly. The  coexistence
of superconductivity   with  antiferromagnetism for 0.35 $\leq$
$x\leq$ 0.5 is observed in the temperature dependence of $1/T_1$
which displays a broad peak at $T_N$ and drops as $T^3$ below
$T_c$. At $x$=0.5, $T_N$ is reduced to 3 K while $T_c$ reaches 0.9
K. Our results  suggest that the coexisting antiferromagnetic
order and superconductivity are due to the same electronic state
derived from the Ce-4f$^1$ electron. It is most interesting that
the superconducting transition temperature $T_c$ is increased as
the system penetrates deeper inside the antiferromagnetically
ordered state. $T_c$ for $x$=0.45 and 0.5 is more than double that
of CeIrIn$_5$. In the coexistence region, $1/T_1$ shows a
$T$-linear dependence at low-$T$ below $T\sim$0.4 K. We have
argued that this may arise from some magnetic excitations
associated with the coexisting magnetism, in addition to the
presence of crystal disorder that produces a residual density of
states at the fermi level.

\section{Acknowledgement}

We thank H. Harima for a helpful discussion on the $\nu_Q$ issue, and  G. G. Lonzarich, N. Nagaosa and S.-C. Zhang for helpful comments. We also would like to thank W. Bao and N.J. Curro for useful discussion, and S. Kawasaki, K. Tanabe and S. Yamaoka for assistance in some of the measurements. Partial support by Japan MEXT grant No. 14540338, 16340104 (G.-q.Z) and No.  10CE2004 (Y.K) is   thanked. Work at Los Alamos was performed under the auspices of the US DOE.

\newpage

* present address: Department of Physics, Okayama University, Okayama 700-8530, Japan. E-mail address: zheng@psun.phys.okayama-u.ac.jp


\begin{references}

\bibitem{Maple&Fisher}
M. B. Maple and O. Fisher (Eds), {\it Superconductivity and Magnetism}, (Springer-Verlag, Berlin, 1982).
\bibitem{Barnhart}
C. Bernhard {\it et al.}, Phys. Rev. {\bf B 61}, R14960 (2000).
\bibitem{Geibel}
C. Geibel {\it et al.}, Z. Phys. {\bf B 84}, 1 (1991).
\bibitem{Tou}
H. Tou {\it et al}, J. Phys. Soc. Jpn. {\bf 64}, 725 (1995).
\bibitem{Sato}
N.~K.~Sato, N.~Aso, K.~Miyake, R.~Shiina, P.~Thalmeier, G.~Varelogiannis, C.~Geibel, F.~Steglich, P.~Fulde and T.~Komatsubara,
Nature {\bf 410}, 340 (2001).
\bibitem{Miyake}
S. Yotsuhashi, H. Kusunose and K. Miyake, J. Phys. Soc. Jpn. {\bf 70}, 186 (2001).
\bibitem{Geibel2}
C. Geibel {\it et al.}, Z. Phys. {\bf B 83}, 305 (1991).
\bibitem{UGe2}
S.~S.~Saxena, P.~Agarwal, K.~Ahilan, F.~M.~Grosche, R.~K.~W.~Haselwimmer, M.~J.~Steiner, E.~Pugh, I.~R.~Walker, S.~R.~Julian, P.~Monthoux, G.~G.~Lonzarich, A.~Huxley, I.~Sheikin, D.~Braithwaite, and J.~Flouquet,
Nature {\bf 406}, 587 (2000).
\bibitem{URhGe}
D.~Aoki, A.~Huxley, E.~Ressouche, D.~Braithwaite, J.~Flouquet, J.~P.~Brison, E.~Lhotel, and C.~Paulsen,
Nature {\bf 413}, 613 (2001).
\bibitem{Zhang}
S.- C. Zhang, Science  {\bf 275}, 1089 (1997).
\bibitem{Pagliuso}
P.G. Pagliuso {\it et al.}, Phys. Rev. {\bf B64}, 100503 (2001).
\bibitem{Maple}
V. S. Zapf, E. J. Freeman, E. D. Bauer, J. Petricka, C. Sirvent, N. A. Frederick, R. P. Dickey, and M. B. Maple, Phys. Rev. {\bf B 65}, 014506 (2002).
\bibitem{Mito}
T. Mito, S. Kawasaki, Y. Kawasaki, G.-q. Zheng, Y. Kitaoka, D Aoki, Y Haga, and Y. Onuki, Phys. Rev. Lett. {\bf 90}, 077004 (2003).
\bibitem{Kawasaki}
S. Kawasaki, T. Mito, Y. Kawasaki, G.-q. Zheng, Y. Kitaoka, D Aoki, Y Haga, and Y. Onuki, Phys. Rev. Lett. {\bf 91}, 137001 (2003).
\bibitem{Hegger}
H.~Hegger, C.~Petrovic, E.~G.~Moshopoulou, M.~F.~Hundley, J.~L.~Sarrao, Z.~Fisk, and J.~D.~Thompson,
Phys.\ Rev.\ Lett. {\bf 84}, 4986 (2000).
\bibitem{Petrovic}
C. Petrovic {\it et al.}, Europhys. Lett. {\bf 53}, 354 (2001).
\bibitem{Zheng}
G.-q. Zheng,  K. Tanabe, T. Mito, S. Kawasaki, Y. Kitaoka, D. Aoki, Y. Haga, and Y.  Onuki,
Phys. Rev. Lett. {\bf 86}, 4664 (2001).
\bibitem{Bao1}
W. Bao {\it et al.}, Phys. Rev.  {\bf B 65}  100505  (2002).

\bibitem{Morris}
G.D. Morris {\em et al} Physica {\bf B 326}, 390 (2003)
\bibitem{ZhengSSC}
G.-q. Zheng, E. Yanase, Y. Kitaoka, K. Asayama, Y. Kodama, R. Tanaka and S. Endo, Solid State Commun. {\bf 79},  51 (1991).
\bibitem{smith}
T. F. Smith, C. W. Chu and M. B. Maple, Cryogenics. {\bf 9}, 53 (1969).
\bibitem{Maclaughlin}
D.E. Maclaughlin {\it al}, Phys. Rev. {\bf B 4}, 60 (1971).
\bibitem{Roos}
J. Chepin and J.H. Ross, J. Phys. Cond. Matt. {\bf 3}, 8103 (1991).
\bibitem{Curro}
N. Curro {\it et al.}, Phys. Rev. {\bf B 62},  6100 (2000).
\bibitem{Harima}
K. Betsuyaku and H. Harima, J. Mag. Mag. Matt. {\bf 272-276}, 187 (2004).
\bibitem{Mito3}
T. Mito, S. Kawasaki, G.-q. Zheng, Y. Kawasaki, K. Ishida, Y. Kitaoka, D. Aoki, Y. Haga,  and Y. Onuki,
Phys. Rev.  {\bf B 63},  220507 (2001).
\bibitem{Kawasaki2}
S. Kawasaki, T. Mito, G.-q. Zheng, C. Thessieu, Y. Kawasaki, K. Ishida, Y. Kitaoka, D. Aoki, S. Araki, Y. Haga, R. Settai and Y. Onuki, Phys. Rev. {\bf B 65}, 020504 (2002).
\bibitem{Mito2}
T. Mito {\em et al}, unpublished.
\bibitem{Kohori}
Y. Kohori {\it et al}, Eur. Phys. J. {\bf B18}, 601 (2000).

\bibitem{Doniach}
S. Doniach,
in {\it Valence Instabilities and Related Narrow
Band Phenomena}, edited by R. D. Parks (Plenum, New
York, 1977), p. 169.
\bibitem{Christianson}
Preliminary neutron results  suggest that the ordered moment for $x$=0.5 is reduced by a factor of 4 from that for the $x$=0 sample  (A. D. Christianson, private communication).
\bibitem{Curro2}
In fact, on going from CeRhIn$_5$ to CeCoIn$_5$, the hyperfine coupling decreases by a factor of 3. See Curro {\em et al}, Cond-mat/0205354.

\bibitem{Miyake2}
S. Schmitt-Rink, K. Miyake and C.M. Varma, Phys. Rev. Lett. {\bf 57}, 2575 (1986); K. Miyake, unpublished data (1991).
\bibitem{Fuseya}
Y. Fuseya, H. Kohno and K. Miyake, J. Phys. Soc.  Jpn. {\bf 72}, 2914 (2003).
\bibitem{Bang}
Y. Bang, M.  J. Graf, A. V. Balatsky and J. D. Thompson, Phys. Rev. {\bf B 69}, 014505 (2004).
\bibitem{Zheng1}
G.-q. Zheng, Y. Kitaoka, K. Asayama, K. Hamada, H. Yamauchi and S. Tanaka, Physica  {\bf C 260}, 197 (1996).


\end{references}
\end{document}